\documentclass[%
 reprint,
nofootinbib,
nobibnotes,
 amsmath,amssymb,
aps,
floatfix,
]{revtex4-1}

\usepackage{graphicx}
\usepackage{dcolumn}
\usepackage{bm}
\usepackage{booktabs}
\usepackage{array}
\usepackage{placeins}

\usepackage[caption=false]{subfig}

\newcommand\blfootnote[1]{%
  \begingroup
  \renewcommand\thefootnote{}\footnote{#1}%
  \addtocounter{footnote}{-1}%
  \endgroup
}

\begin{document}


\title{Pressure-induced suppression of ferromagnetism in the itinerant ferromagnet LaCrSb$_3$}

\author{Z.~E. Brubaker$^{1,2,3}$, J.~S. Harvey$^{2}$, J.~R. Badger$^{2}$, R.~R. Ullah$^{2}$, D.~J. Campbell$^{4}$, Y. Xiao$^{5}$, P. Chow$^{5}$, C. Kenney-Benson$^{5}$, J.~S. Smith $^{5}$, C. Reynolds$^{3}$, J. Paglione$^{4}$, R.~J. Zieve$^{2}$, J.~R. Jeffries$^{3}$, and V. Taufour$^{2}$\\
$^{1}$ Oak Ridge National Laboratory, Oak Ridge, Tennessee 37831, USA \\
$^{2}$ University of California, Davis, California 95616, USA\\
$^{3}$ Lawrence Livermore National Laboratory, Livermore, California 94550, USA\\
$^{4}$ Maryland Quantum Materials Center, Department of Physics, University of Maryland, College Park, MD 20742, USA\\
$^{5}$ HP-CAT, X-ray Science Division, Argonne National Laboratory, Lemont, Illinois 60439, USA\\
}
\date{\today}

\begin{abstract}


We have performed an extensive pressure-dependent structural, spectroscopic, and electrical transport study of LaCrSb$_3$. The ferromagnetic phase (T$_C$~= 120~K at p~= 0 GPa) is fully suppressed by p~= 26.5~GPa and the Cr-moment decreases steadily with increasing pressure. The unit cell volume decreases smoothly up to p~= 55~GPa. We find that the bulk modulus and suppression of the magnetism are in good agreement with theoretical predictions, but the Cr-moment decreases smoothly with pressure, in contrast to steplike drops predicted by theory. The ferromagnetic ordering temperature appears to be driven by the Cr-moment.

\end{abstract}
\pacs{Valid PACS appear here}
\maketitle


\section{Introduction}

\blfootnote{Notice:  This manuscript has been authored in part by UT-Battelle, LLC, under contract DE-AC05-00OR22725 with the US Department of Energy (DOE). The US government retains and the publisher, by accepting the article for publication, acknowledges that the US government retains a nonexclusive, paid-up, irrevocable, worldwide license to publish or reproduce the published form of this manuscript, or allow others to do so, for US government purposes. DOE will provide public access to these results of federally sponsored research in accordance with the DOE Public Access Plan (http://energy.gov/downloads/doe-public-access-plan).}

The features of second-order quantum phase transitions have been explored experimentally for nearly half a century, and general theories in recent decades have made great strides in describing the myriad features observed near these quantum critical points (QCP): exotic superconductivity, non-Fermi liquid behavior, the emergence of heavy Fermion quasiparticles, spin resonances, etc. \cite{Hertz, Millis, Lohneysen}. Although antiferromagnetic ordering driven toward zero temperature often shows the hallmarks of second-order QCP, ferromagnetic transitions similarly driven toward zero temperature exhibit different emergent properties. The ferromagnetic states rarely suppress continuously to zero temperature, instead showing first-order transitions in the absence of magnetic fields \cite{Belitz,Brando}. How ferromagnetic order yields near these transitions is an active area of research, but there are few materials that have been tuned toward criticality and even fewer experiments that probe the relevant physical properties of those materials as a function of tuning parameter. 

The ferromagnet LaCrSb$_3$ is a particularly promising candidate to investigate ferromagnetic criticality tuned toward zero temperature, because density functional theory calculations predict the pressure-induced suppression of the Cr-moment in an unusual set of abrupt steps \cite{Nguyen}. If the theoretical predictions regarding the Cr-moment can be confirmed, then this would establish the magnetic state of LaCrSb$_3$ near the critical region, providing a firm footing to potentially understand the properties in that region. 

LaCrSb$_3$ orders in the orthorhombic Pbcm structure with lattice constants of a~= 13.266~\AA{}, b~= 6.188~\AA{}, and c~= 6.100~\AA{}, with the nearest neighbor Cr--Cr spacing being one-half of the c-axis. The ferromagnetic transition at ambient pressure occurs between 125 and 142~K depending on the sample and determination method of T$_C$ and is readily discernible in both resistivity and magnetization measurements \cite{Raju,Hartjes,Leonard_1,Leonard_2}. A spin reorientation in the \textit{bc}-plane near 95~K also exists, which can be suppressed with a small field of approximately 250~Oe \cite{Raju, Granado}. The magnetic structure is nontrivial and has been studied with neutron diffraction, $\mu$SR, x-ray photoelectron spectroscopy, as well as various theoretical models, which have shown coexisting ferromagnetic and antiferromagnetic sublattices below T$_C$ \cite{Granado, MacFarlane, Crerar, Choi, Nguyen}. Chemical substitution on the Cr site with V, Mn, Fe, Cu, and Al has been performed and shows a general suppression of the ordering temperature with substitution \cite{Magnetization, Dubenko}. High-pressure magnetization measurements up to p~= 6~GPa, however, have shown that the ferromagnetic transition is robust against these pressures, though the spin reorientation is suppressed by about 3~GPa \cite{Magnetization}.
 
\begin{figure*}[!htb]
	\centering
	\includegraphics[width=\linewidth]{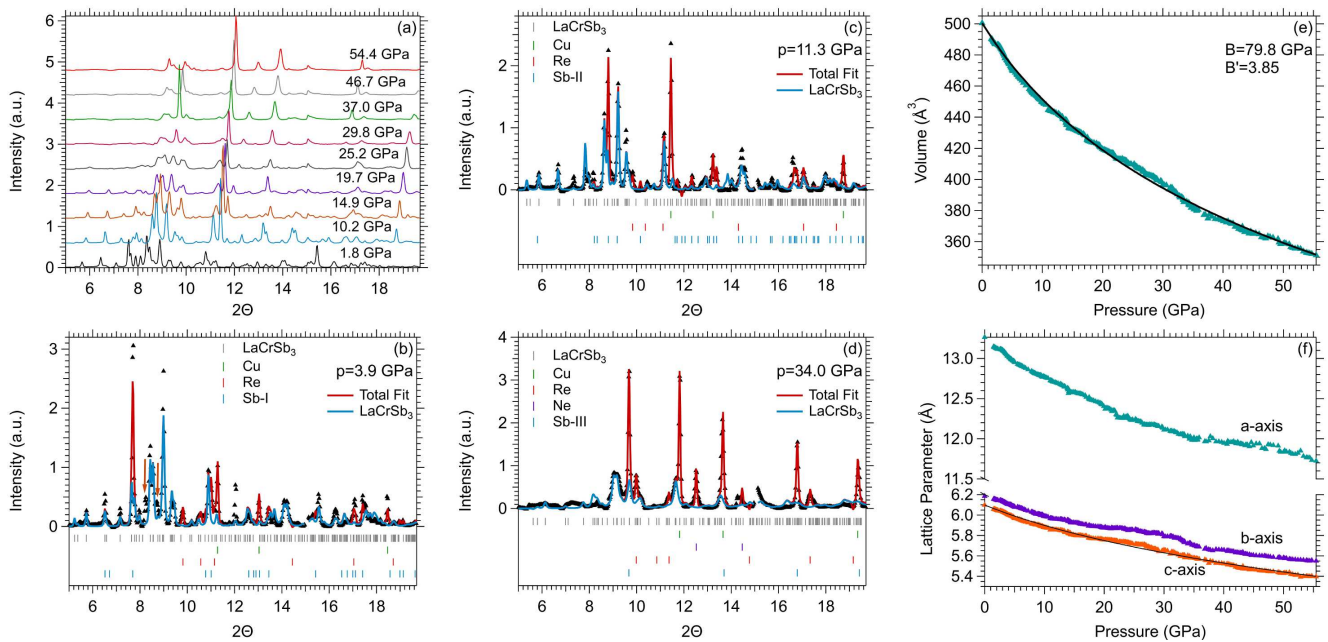}
	\caption{(a) PXRD spectra at select pressures for LaCrSb$_3$. (b-d) Total fit and LaCrSb$_3$ contribution for spectra containing Sb-I, Sb-II, and Sb-III. Peak positions for each phase are indicated; from top to bottom the peak locations correspond to LaCrSb$_3$, Cu, Re, Sb-I/Sb-II in (b) and (c) and to LaCrSb$_3$, Cu, Re, Ne, and Sb-III in (d). The small jump observed near 15 degrees in (a) was excluded in the fitting, which is why there is a small gap in (b-d) near 15 degrees. The arrows in (b) near 8 degrees indicate unindexed peaks which disappeared by about p=10 GPa. Unlike LaCrSb$_3$, these did not show any single crystal peaks in the 2-D pattern, making these unlikely to be related to LaCrSb$_3$. Overall, the LaCrSb$_3$ peaks broaden considerably and decrease in intensity under pressure. (e) Evolution of volume with pressure. The solid line is a fit to the Birch--Murnaghan equation of state, which yields values of B~= 79.8~GPa and B$^\prime$~= 3.85 in good agreement with theoretical predictions \cite{Nguyen}. (f) Lattice parameters found from the PXRD spectra. The solid line is a fit to the Birch--Murnaghan equation of state for the c-axis data (performed by fitting to V=c$^3$) and is used to convert pressure to Cr--Cr distance. }
	\label{fig:PXRD_spectra}
\end{figure*}  
 
The experimental characterization of materials under high pressures are often limited to a few properties, which is usually insufficient for a detailed understanding of the physics. We recently showed that band-structure calculations and experiments can be used cooperatively to identify compounds with fragile magnetic states \cite{Nguyen}. Here, we assess the Curie temperature, crystal structure, and evolution of magnetic moment under pressure by combining transport, x-ray diffraction, and x-ray emission measurements. We find that the bulk modulus and suppression of the magnetism are in good agreement with theory, though the Cr-moment is suppressed smoothly, in contrast to predictions.

\section{Experimental Methods}

LaCrSb$_3$ was grown using an Sb-flux following the parameters used by Lin and colleagues \cite{Magnetization}. Powder x-ray diffraction (PXRD) measurements were performed at sector 16-IDB of the Advanced Photon Source using a 30~keV incident beam. Powdered samples were loaded into a rhenium gasket in a diamond anvil cell (DAC) along with copper powder to calibrate the pressure and were gas-loaded with neon. The gasket was pre-indented to 40 $\mu$m using diamonds with 300 $\mu$m culets, and a 130 $\mu$m hole was drilled using an electric discharge machine. Pressure was controlled with a gas-driven membrane. The acquisition time was 2~seconds for pressures below 33~GPa and 10~seconds for pressures above 33~GPa to account for the intensity loss with increasing pressure. All analysis was performed using Fit2D and GSAS-II \cite{Fit2D,GSAS2}. The relative pressure gradient for PXRD measurements was likely smaller than the pressure gradient in x-ray emission and transport measurements because Ne was used as the pressure medium.

\begin{figure*}[phtb]
	\centering
	\includegraphics[width=\linewidth]{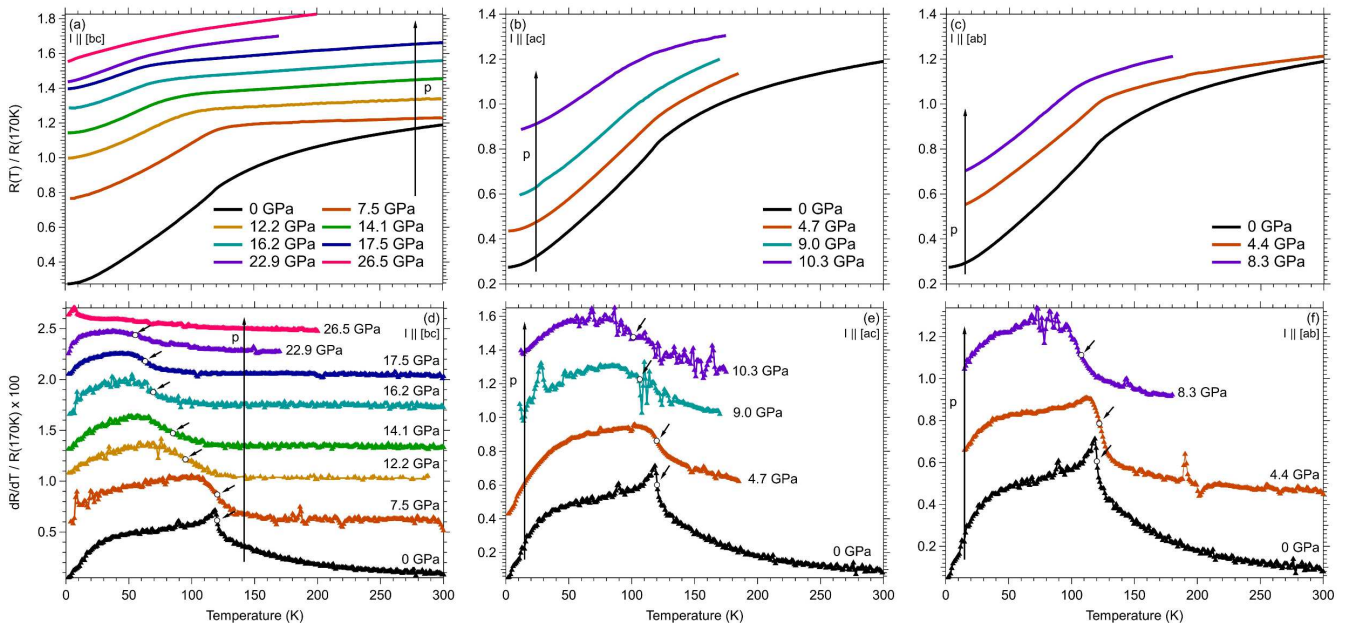}
	\caption{(a--c) Resistivity measurements and (d--f) derivatives with the current and voltage in (a,d) the [bc] plane, (b,e) the [ac] plane, and (c,f) the [ab] plane. Each plot includes the same P=0~GPa resistivity measurement; this was not repeated for different current directions. For the 7.5~GPa and 12.2~GPa data in (d) we only plot the derivative for data taken with constant gain on the PPMS resistance bridge, resulting in a lower data point density. Above these pressures, the gain was held constant. The arrows and white circles indicate T$_C$. The transition width corresponds to the linear region of dR/dT / R(170 K) near T$_C$, and T$_C$ corresponds to the midpoint of the transition width, which is roughly equivalent to the inflection point. Several curves in (d-f) show sharp jumps below 10 K and also near 150-200 K, which were irreproducible and we attribute them to bad electrical contacts.}
	\label{fig:transport}
\end{figure*}

X-ray emission spectroscopy (XES) was performed at sector 16-IDD with a constant incident energy of 11.6~keV, well above the Cr K-edge just below 6~keV. A small crystal was loaded into a beryllium gasket, along with silicone oil as pressure transmitting medium and a ruby. The beryllium gasket was pre-indented to a thickness of 40 $\mu$m using diamonds with 200 $\mu$m culet sizes and a 45 $\mu$m hole was drilled using a laser drilling system available at HPCAT. Pressure was controlled with a gas-driven membrane and checked with ruby fluorescence \cite{Mao}. Both PXRD and XES measurements were performed at T~= 300~K. Each XES scan was performed from 5907~eV to 5972~eV using an energy spacing of 0.4~eV and an exposure time of 12~seconds. These scans were repeated and summed at each pressure until the uncertainty due to counting statistics at the K-beta peak was less than 1.5\%. As a result, the lowest measured pressures only required two to three repetitions, but the highest measured pressures required up to nine iterations to obtain satisfactory spectra. Because only one ruby was measured, we could not determine the pressure gradient, but based on similar experiments we have performed at 16-IDD, we estimate the pressure gradient to be about 1 to 2 GPa at the highest measured pressures.

Transport measurements were performed using a Be--Cu designer diamond anvil cell (DDAC) designed to fit inside the Quantum Design physical property measurement system (PPMS). A standard 300-$\mu$m culet was paired with a 250-$\mu$m designer diamond culet with eight tungsten probes lithographically deposited onto the culet and encapsulated with deposited synthetic diamond \cite{MiniDAC_1,MiniDAC_2,MiniDAC_3}. We used a nonmagnetic MP35N gasket, which was indented to 40~$\mu$m and an 80-$\mu$m hole was drilled using an electric discharge machine. Samples with sizes of approximately $50 \times 50 \times 10$~$\mu$m were placed on the tungsten probes. Steatite was used as a quasi-hydrostatic pressure transmitting medium, and one to two rubies were loaded to measure the pressure. In instances when multiple rubies were measured, the pressures of the two rubies were averaged; at low pressures the rubies agreed to within 0.5~GPa, but at the highest measured pressures a difference of approximately 2 to 3 GPa was observed. The DDAC is designed to minimize pressurization of the piston due to differential thermal contraction effects and the pressure variation with temperature is estimated to be about 5\%. \cite{Jackson} The applied magnetic field is along the thin direction, and the current and voltage paths are perpendicular to this direction. To change orientation a newly prepared crystal must be mounted in the DDAC.

\begin{figure*}[!htpb]
	\centering
	\includegraphics[width=\linewidth]{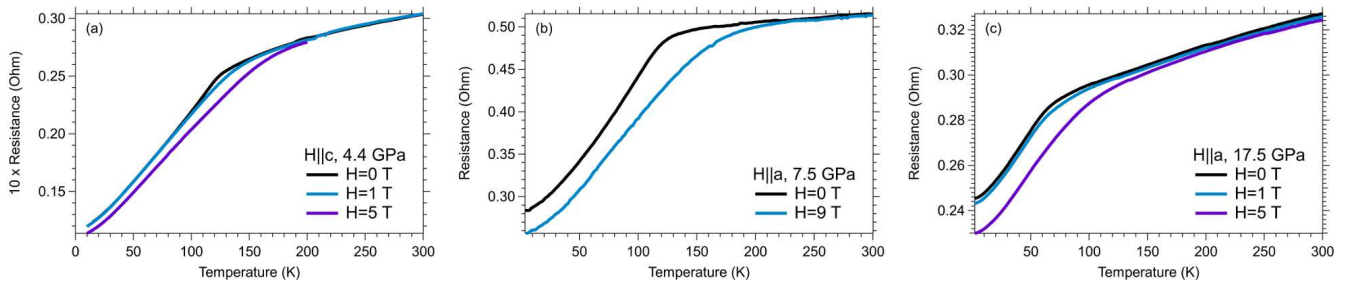}
	\caption{Constant field resistivity measurements at select pressures with field along the (a) c-axis and (b--c) a-axis.}
	\label{fig:withField}
\end{figure*}

The LaCrSb$_3$ crystals cleave perpendicular to the a-axis providing the most robust samples with B$||$a and I$||$[bc]. For measurements with B$||$b (I$||$[ac]) and B$||$c (I$||$[ab]) samples were oriented to the appropriate direction, then polished to 10-$\mu$m thickness. The polished samples are far more sensitive to pressure gradients and easily crack along the a-axis. Because of this, the crystals with I$||$[ac] and I$||$[ab] failed at far lower pressures than the crystals with I$||$[bc] and in general are more prone to glitches in the data.

All resistivity data was taken on warming. The I$||$[bc] measurements below 14~GPa used a rate of r~= 1.0--1.5~K/min, an excitation current of I~= 0.1~mA, and allowed the gain setting to vary. All other measurements were performed with a sweep rate of r~= 0.2~K/min below 200~K and r~= 1~K/min from 200~K to 300~K, with an excitation current of 1~mA, and a fixed gain of either 0.400~$\mu$V or 1.0~mV. The fixed gain was necessary to prevent slight differences that were observed in lower-pressure measurements between varying gain settings. 

For all pressure measurements, the average of the pressure before and after each measurement is presented herein; the errorbars correspond to the pressure before and after each measurement.

\section{Results}
\subsection{Structural Studies}

Figure~\ref{fig:PXRD_spectra}a shows representative PXRD patterns, which were collected to just below 55~GPa. The peaks broaden considerably under pressure and many of the low-intensity peaks visible at 1.8~GPa disappear by 20--30~GPa. In addition to LaCrSb$_3$, the diffraction patterns show evidence of Cu (pressure marker), Ne (pressure medium), Re (from the gasket), and Sb (used in the flux growth). Antimony undergoes several structural transitions---(i)~rhombohedral (Sb-I)-host-guest (HG) monoclinic (Sb-IV) at 8.2~GPa, (ii)~HG monoclinic-HG tetragonal (Sb-II) at 9~GPa, and (iii)~HG tetragonal-cubic (Sb-III) at 28~GPa---which need to be accounted for to obtain values for the lattice parameters of LaCrSb$_3$ \cite{Sb, Sb_2}. In the collected data, we did not observe the Sb-IV phase, though the narrow pressure range and subtle differences in the diffraction patterns between Sb-IV and Sb-II make this unsurprising.

Figures~\ref{fig:PXRD_spectra}b-d show three representative fits for patterns in which Sb-I, Sb-II, and Sb-III are observed. The contribution of LaCrSb$_3$ is explicitly shown in these figures, as well as the diffraction peak locations of each phase. The location of the calculated diffraction peaks agree well with the observed data, though not all of the intensities are satisfactorily fit. This is largely due to strong single crystal peaks observed in the 2-D patterns. The fit quality can be improved by including a large number of preferred orientation parameters. Although this improves the overall fit quality, the varying degree of single crystal peaks in each diffraction pattern introduces quite a bit of noise in the extracted volume and lattice parameters under pressure depending on exactly which single crystal peaks are present. For this reason, we limited the preferred  orientation parameters to a harmonic order of four, which yielded the best combination of fit quality and consistent pressure-dependent structural parameters. 

Figures~\ref{fig:PXRD_spectra}e-f show the obtained volume and lattice parameters under pressure. The volume can be fit with the Birch--Murnaghan equation of state and yields values of 79.8~GPa and 3.85 for the bulk modulus (B) and its derivative (B$^\prime$) respectively, in good agreement with theoretical predictions of B~= 78~GPa \cite{BM,Nguyen}. The diffraction patterns do not show any new diffraction peaks emerging under pressure, implying that no structural transition to a different Bravais lattice or space group occurs under pressure up to 55~GPa. The varying sample intensities due to single crystal peaks, however, make it impossible to determine if the atomic positions are rearranged. Nonetheless, all analysis of the diffraction patterns was performed assuming the atomic positions remain unchanged under pressure. This assumption allows for an estimate of the nearest neighbor Cr--Cr distance, which is one-half of the c-axis at ambient pressure. To provide a smooth Cr--Cr distance curve under pressure, we have fit the c-axis with the Birch--Murnaghan equation of state as shown in Fig.~\ref{fig:PXRD_spectra}f, yielding B~= 84.7~GPa and B$^\prime$~= 3.3. This function has been used for all subsequent conversions from pressure to Cr--Cr distance.

We note that the volume curve shows a slight deviation from the Birch-Murnaghan equation of state around 25-30 GPa, which is caused by the apparent stiffening of the b- and c-axes in this pressure range. It is unclear if this deviation is real, or an artifact of the fitting (perhaps due to the overlapping Sb peaks), though the fact that the data can be fit to the equation of state satisfactorily below and above this pressure range suggests that this is caused by the fitting.

\begin{figure}[htbp]
	\centering
	\includegraphics[width=\linewidth]{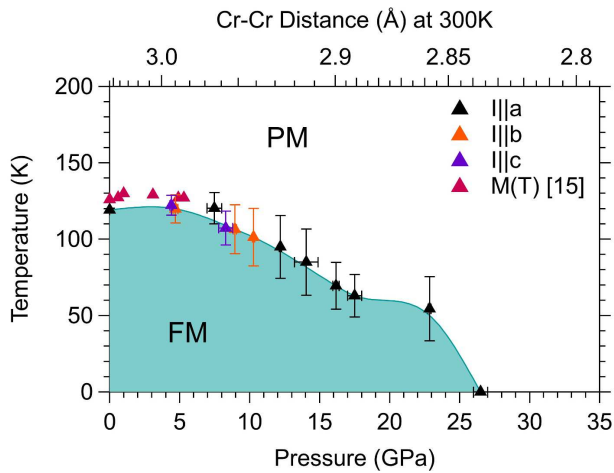}
	\caption{Phase diagram determined from resistivity (this work) and magnetization measurements \cite{Magnetization}. The ferromagnetic phase is robust against pressures below 6~GPa and shows a nearly linear suppression beyond 6~GPa. Our data also suggests a plateau in T$_C$ from 17.5 to 23~GPa and a sudden suppression at the highest measured pressures, though this hinges on a single data point. Error bars correspond to the transition width, as determined by the linear region of dR/dT / R(170 K) near T$_C$.}
	\label{fig:PhaseDiag}
\end{figure}

\begin{figure*}[!htbp]
	\centering
	\includegraphics[width=\linewidth]{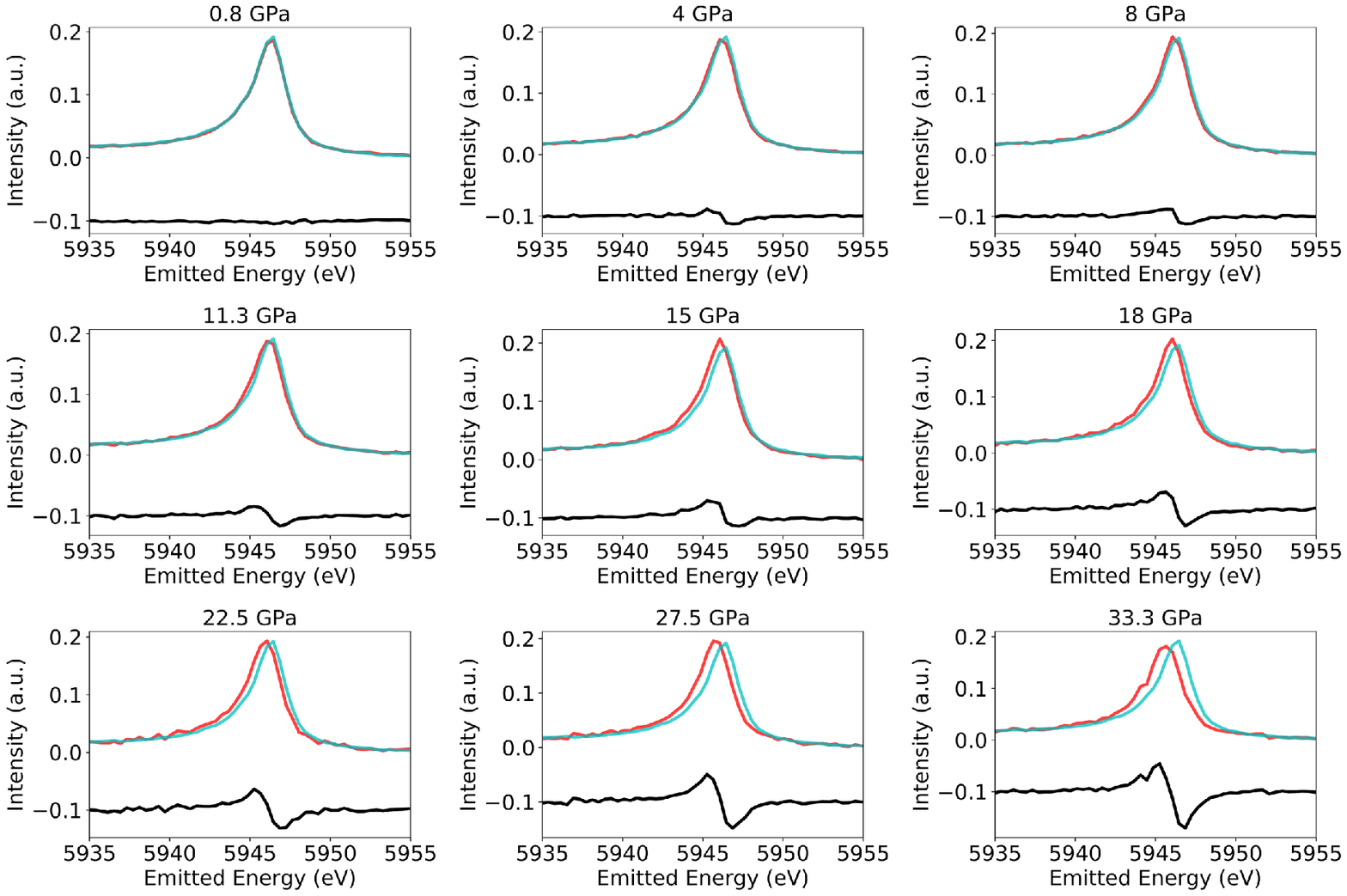}
	\caption{X-ray emission spectra for each of the measured pressure. The cyan curves are the ambient pressure measurement, the red curves are the pressurized measurements, and the black curve is the difference between the two curves and is shifted by $-0.1$. The K$\beta$ peak gradually shifts to lower energies, indicating a decreasing Cr-moment.}
	\label{fig:XES_spectra}
\end{figure*}

\subsection{Transport Measurements}

Resistivity measurements were performed up to p~= 26.5~GPa with I$||$[bc], up to p~= 10.3~GPa with I$||$[ac], and up to p~= 7.8~GPa with I$||$[ab]. The resistivity measurements are summarized in Fig.~\ref{fig:transport}, and the field-dependent measurements are shown in Fig.~\ref{fig:withField}. Ambient pressure measurements show a kink in resistivity at T~= 120~K, which is a signature of the PM--FM transition observed in LaCrSb$_3$, and is close to the ordering temperature reported in the literature. Under pressure, the kink broadens into a gradual change in slope and moves to lower temperatures, suggesting a gradual suppression of the FM transition with pressure. By p~= 26.5~GPa there no longer exists a discernible signature of the PM--FM transition. The ferromagnetic nature of the observed transition is confirmed by the fact that it rapidly broadens under magnetic field (Fig.~\ref{fig:withField}), which is expected because the time-reversal symmetry is broken by the field. As was done with LaCrGe$_3$ and UGe$_2$, we define the midpoint of the change in slope as T$_C$, which is shown with a white circle and arrow in Figs.~\ref{fig:transport}d--f \cite{LaCrGe3, UGe2}. We note that relatively sharp anomalies can be observed on various curves below about 10~K and near 150-200~K. These anomalous features were irreproducible, and we attribute them to bad electrical contacts on the tiny samples, which manifest at low temperatures due to thermal contraction effects. These issues were exacerbated with increasing magnetic field.

The temperature-pressure phase diagram can be constructed by combining our results with the previously published low-pressure magnetization measurements and is shown in Fig.~\ref{fig:PhaseDiag} \cite{Magnetization}. The ordering temperature remains robust against pressures up to about 6~GPa, at which point the ferromagnetic phase is nearly linearly suppressed. Our results also suggest a plateau from 17.5 to 23~GPa followed by an abrupt suppression of the FM phase, though this hinges on a single data point shortly before the sample broke. If this can be confirmed, the sudden suppression of the FM phase at high pressure may be an indication that the transition has become first-order as expected in clean metallic ferromagnets \cite{Belitz,Brando}.

\subsection{X-ray Emission Spectroscopy}

Using x-ray emission spectroscopy to quantify the local spin of a system has successfully been used for various transition metal compounds \cite{Fe_XES, Fe_jeffries, FeSe, CoOxide,CoOxide_2, Mn_XES,MnOxide}. The indirect K$\beta$ fluorescence (3p$\rightarrow$ 1s) is split into a main K$\beta _{1,3}$ peak and a broad K$\beta ^\prime$ shoulder because of the exchange interaction between the spins of the transition metal 3$d$ states and the spin of the unpaired p-electron after the K$\beta$ transition \cite{XES_spin}. A change in the local spin will then manifest itself as a change in both peak locations and intensities \cite{IAD,HardXES,Cr_XES}. To quantify the moment, collecting a reference spectrum to which compare subsequent spectra is necessary.

To analyze the XES spectra, we used the integrated absolute difference method (IAD) \cite{IAD}. This method is considered more reliable than simply using the energy and intensity shift to approximate the local moment \cite{HardXES}. In the IAD method, the absolute difference of two spectra is integrated to determine the relative difference between them. Specifically, 

\begin{equation}
IAD_i = \int_{E_1}^{E_2}|\sigma_i(E) - \sigma_0(E)|dE ,
\end{equation}

where $\sigma_0$ is the reference spectrum and $\sigma_i$ is the investigated spectrum. The IAD value is proportional to the spin of the system. (see Fig.~\ref{fig:XES_spectra}a). 

\begin{figure}[!htbp]
	\centering
	\includegraphics[width=\linewidth]{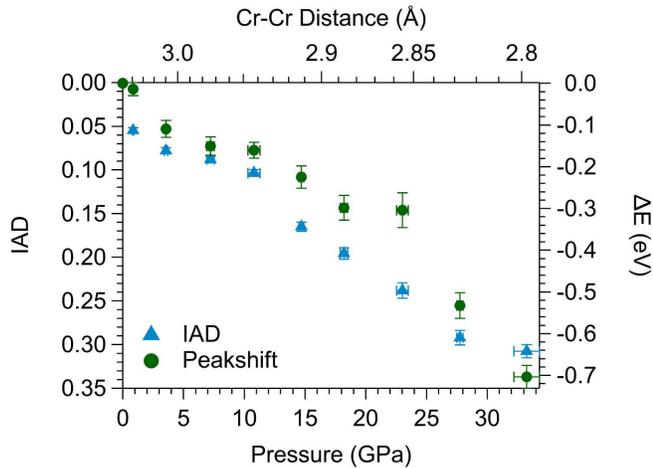}
	\caption{IAD values (blue triangles) and peakshift (green circles) plotted vs. pressure (bottom) and Cr--Cr distance (top). Both IAD values and the peakshift indicate similar qualitative behavior.}
	\label{fig:IAD}
\end{figure}

Figure~\ref{fig:XES_spectra} shows the collected emission spectra from 5935 eV to 5955 eV for each of the measured pressures, the ambient pressure reference spectrum, and the difference between the curves. A steady shift in the primary K$\beta$ peak is observed, reaching a maximum of $\Delta$E~= 0.7~eV at 33~GPa. 
To provide a quantitative description of the Cr-moment, we have calculated the IAD values for each of the measured spectra over the entire measured energy range, E$_e$~= 5907~eV to E$_e$~= 5972~eV. The resulting IAD values are shown in Fig.~\ref{fig:IAD}. Overall, the IAD value (i)~increases modestly to about 10~GPa,  (ii)~increases quickly from 10~GPa to 26.5~GPa, and (iii)~levels off above 26.5~GPa. We point out that the IAD value at p=0.8~GPa is due to fluctuations and noise in the collected spectra at 0.8~GPa --- as evidenced by the difference curve, the peak has not begun to shift. 

\begin{figure}[!htbp]
	\centering
	\includegraphics[width=\linewidth]{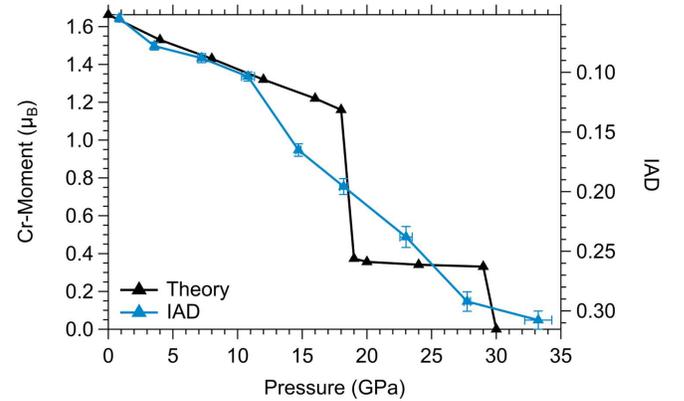}
	\caption{Comparison of experimental IAD values with the predicted pressure dependence of the Cr-moment \cite{Nguyen}. The low pressure region agrees well, but our measurements show a steady decline in the Cr-moment rather than two steplike features.}
	\label{fig:Theory_comp}
\end{figure}

\begin{figure}[!htbp]
	\centering
	\includegraphics[width=\linewidth]{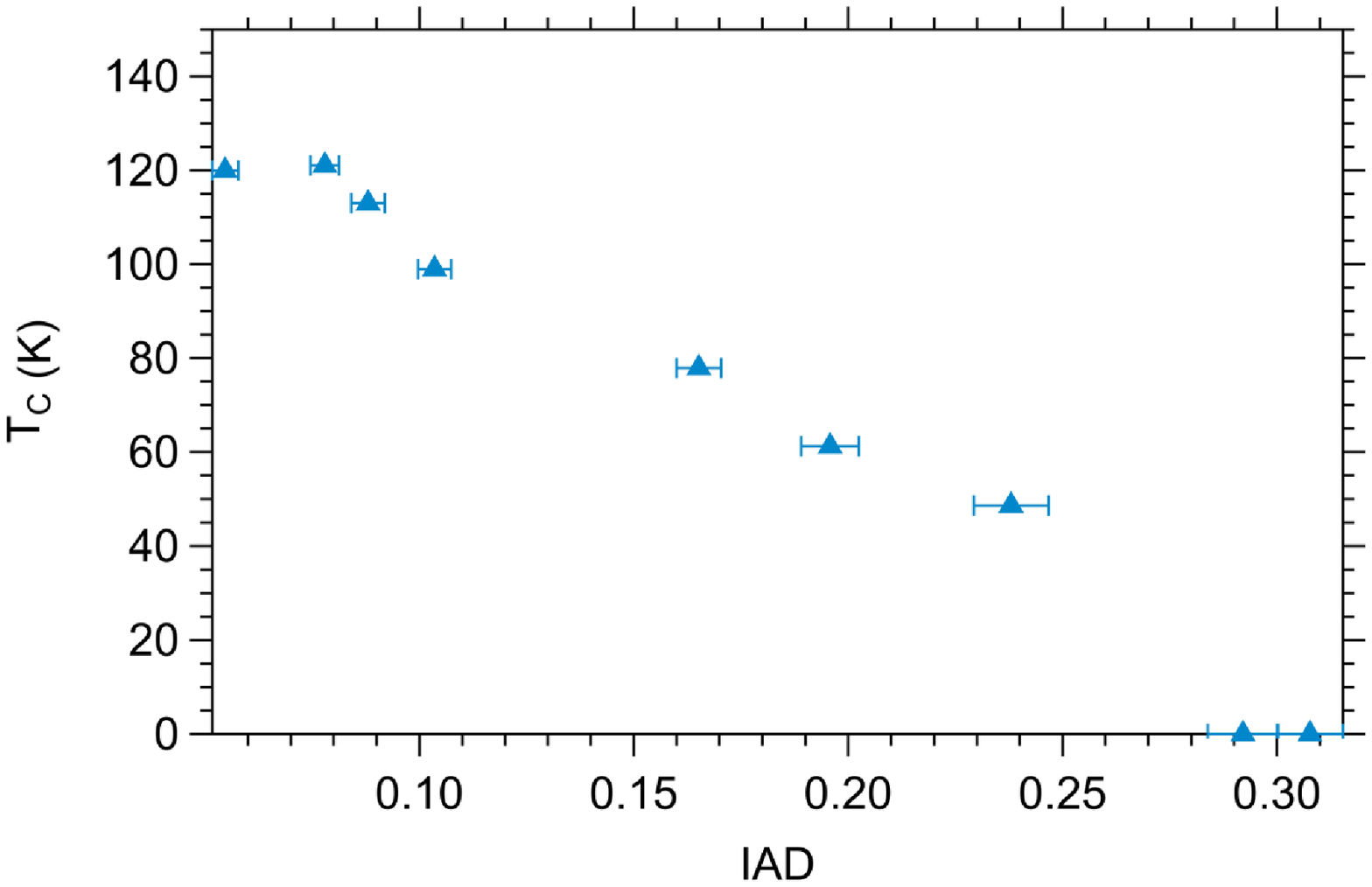}
	\caption{Ordering temperature plotted against IAD value. Error bars for T$_C$ have been omitted for clarity.}
	\label{fig:TCIAD}
\end{figure}

\section{Discussion}

The obtained IAD values can be qualitatively compared to theoretical calculations, as shown in Fig.~\ref{fig:Theory_comp} \cite{Nguyen}. The low-pressure slope agrees well between the experimental and theoretical data, though the IAD values suggest that the Cr-moment steadily decreases from about p~= 11~GPa up until the highest measured pressures, whereas the theoretical model predicts two abrupt steps. This may be due to the experiment being performed at T~= 300~K rather than cryogenic temperatures, which could feasibly broaden these transitions. However, work performed on Co and Fe compounds at both cryogenic temperatures and room temperature demonstrated modest changes in IAD values and slope and did not result in any sharp or steplike features \cite{Fe_jeffries, CoOxide}. Additionally, it would not be unreasonable to expect the phase diagram to reflect a sudden drop in the Cr-moment, which is certainly not the case in our data. The only region of the phase diagram where a steplike feature is possible is at the highest measured pressures when the ferromagnetic phase is fully suppressed. Indeed, a first-order transition is even expected for a ferromagnet approaching a quantum critical point \cite{Belitz, Brando}.  There is, however, no evidence of any sudden change at lower pressures, in contrast to the predictions of the Cr-moment. These considerations suggest that the Cr-moment is, in fact, gradually suppressed up to the highest measured pressures and that this is not an artifact of the temperature at which this experiment was performed. This also implies that the Cr d-orbitals with m=$\pm$2 and m=$\pm$1 show broader peaks in the density of states near the Fermi energy than predicted by calculations, because sharp peaks produce abrupt features like the predicted steplike drops in Cr-moment \cite{Nguyen}.

Figure~\ref{fig:TCIAD} shows the interpolated ordering temperature as a function of IAD value. Plotted in this way, it becomes clear that the ordering temperature is correlated with the IAD value and, thus, the Cr-moment. The correlation of the ordering temperature and the Cr-moment suggests some degree of itinerancy in this system, though V-substitution experiments support a more local picture due to the robustness of the Cr-moment against V-substitution \cite{Magnetization}. In fact, neutron diffraction experiments have suggested that LaCrSb$_3$ falls somewhere between a fully itinerant and local system, which may explain the differences observed between pressure and chemical substitution experiments \cite{Granado}. Our results are consistent with the conclusions of the neutron diffraction studies and suggest that LaCrSb$_3$ cannot be understood in an exclusively local or itinerant picture.

\section{Conclusion}

We have fully suppressed the ferromagnetic phase in LaCrSb$_3$ at p~= 26.5~GPa, which occurs in the absence of any structural phase transition. The ferromagnetic ordering temperature appears to be driven by the Cr-moment. Our results are consistent with an itinerant ferromagnet with the degree of itinerancy increasing with pressure. While the pressure dependence of the Cr-moment agrees well with theoretical predictions at low pressures, we find a gradual suppression of the Cr-moment rather than the predicted pair of steplike features.

\section{Acknowledgments}
The authors thank Feng Zhang and Kai-Ming Ho for fruitful discussion. This work was performed under LDRD (Tracking Code 18-SI-001) and under the auspices of the US Department of Energy by Lawrence Livermore National Laboratory (LLNL) under Contract No. DE-AC52- 07NA27344. Partial funding was provided through the LLNL Livermore Graduate Scholar Program. Portions of this work were performed at HPCAT (Sector 16), Advanced Photon Source (APS), Argonne National Laboratory. HPCAT operations are supported by DOE-NNSA’s Office of Experimental Sciences. The Advanced Photon Source is a U.S. Department of Energy (DOE) Office of Science User Facility operated for the DOE Office of Science by Argonne National Laboratory under Contract No. DE-AC02-06CH11357. This material is based upon work supported by the National Science Foundation under Grant No. NSF DMR-1609855. D.J.C acknowledges the support of the U.S. Department of Energy, Office of Science, Office of Workforce Development for Teachers and Scientists, Office of Science Graduate Student Research program, administered by the Oak Ridge Institute for Science and Education for the DOE under contract no. DE‐SC0014664. J.P. and D.J.C. acknowledge support from the National Science Foundation grant no. DMR-1905891, and the 
Gordon and Betty Moore Foundation's EPiQS Initiative through grant no. GBMF9071

\FloatBarrier

\end{document}